\documentstyle[sprocl,epsfig]{article}

\def\be{\begin{equation}}
\def\ee{\end{equation}}
\def\bea{\begin{eqnarray}}
\def\eea{\end{eqnarray}}

\begin{document}

\title{EeV Neutrinos}

\author{Jaime Alvarez-Mu\~niz and Enrique Zas}

\address{Dept. de F\'\i sica de Part\'\i culas, Universidade de 
Santiago de Compostela\\
E-15706 Santiago de Compostela, Spain\\E-mail: jaime@fpaxp2.usc.es, 
zas@gaes.usc.es} 




\maketitle\abstracts{We discuss the recent developments in the study 
of alternative detection possibilities offered by the radio technique 
and air shower arrays.}


Neutrinos are one of the least explored fundamental sectors of the 
Standard Model because they have a challenging low cross section. 
Besides man made nuclear reactions and beam dump experiments in 
accelerators, neutrinos have only been detected from two astrophysical 
sources: the Sun \cite{solar} and supernova SN 1987A \cite{sn1987A} and from 
the interactions of cosmic rays with the atmosphere \cite{kamioka}. 
Their potential for fundamental research is however large as 
illustrated by the few events 
from SN 1987A \cite{SNnumass} and the recent evidence for flavor 
oscillations \cite{oscillations}, completing the Standard Model and 
providing clues for physics beyond. Moreover they have an unique 
Astronomy potential as they 
can travel unattenuated through matter shields that are opaque to 
other types of radiation. 

The neutrino nucleon cross section rises with energy, first linearly 
and then more slowly because of the low $x$ behavior of the parton 
distributions, so that the Earth becomes opaque for neutrinos of 
$E_{\nu} \sim 100~$TeV, with relevant implications for 
detection techniques. In this article we will concentrate 
on neutrinos above the EeV energy scale with an expected cross 
section in the 10-100~$nb$ range.  

Existing neutrino detectors as well as those in construction or 
planning have motivated estimates of neutrino fluxes from many 
possible sources such as Active Galactic Nuclei (AGN) cores 
\cite{stecore} and jets \cite{jet}, 
Gamma Ray Bursts (GRB) \cite{waxmangrb} and decays of Topological 
Defects (TD) \cite{TD}. These 
calculations extend to energies in the EeV range with fluxes 
that are however quite uncertain because AGN and GRB 
are not well understood and the TD densities and annihilation 
rates are quite unknown. 
There are however better established neutrino fluxes from beam dumps 
in which cosmic rays interact with matter in 
the Universe, either the galactic disk, 
molecular clouds or the Earth atmosphere. Below 100~TeV atmospheric 
neutrinos are subject to uncertainties in the $20\%$ range \cite{physrep} 
but above these uncertainties become larger because prompt decays from 
charm production dominates. 
The establishment of cosmic rays above 
the Greisen-Zatsepin-Kuz'min (GZK) cutoff ($\sim 6~10^{19}$~eV) 
\cite{HECR} and the absorption of protons and nuclei in the Cosmic 
Microwave Background (CMB), guarantees neutrinos of very high 
energies, provided these cosmic rays are of extragalactic origin, 
as most commonly believed. In some models neutrinos of energies 
above $10^{\sim 20}$~eV act as ''messengers'' interacting with the 
cosmic neutrino background to produce cosmic rays \cite{weiler}.

\begin{figure}[hbt]
\centering
\mbox{\epsfig{figure=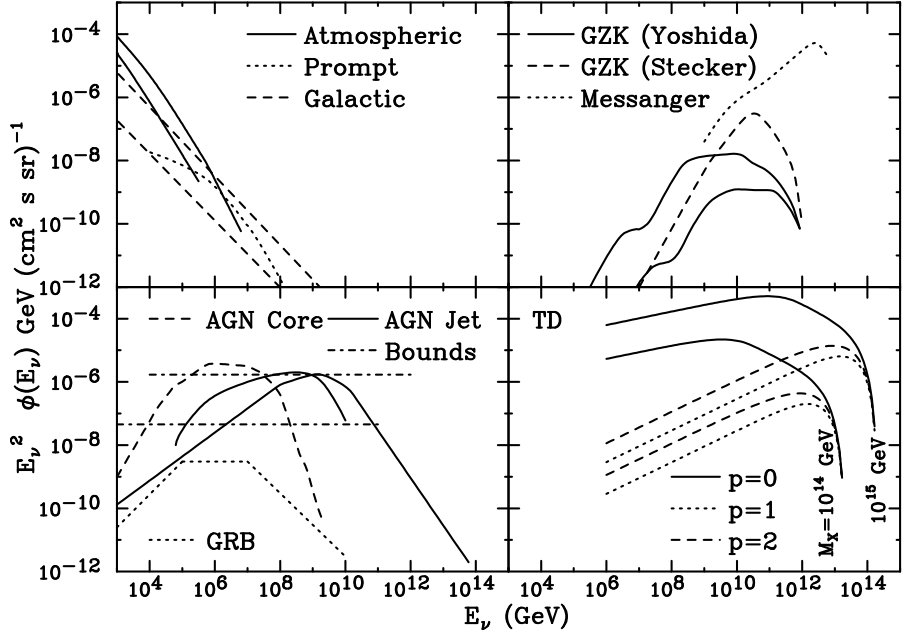,height=8.5cm}}
\caption{Neutrino flux predictions as labelled and bounds for optically 
thin (lower) and thick (upper) photon targets, see text. }
\label{fluxes}
\end{figure}

Fig. \ref{fluxes} compares the atmospheric flux, to calculations of: 
prompt neutrinos \cite{Volkova}, those from cosmic ray interactions 
with galactic matter \cite{domokos} and with the CMB \cite{nuGZK}, 
a messenger neutrino model \cite{sigl}, production in AGN jets 
\cite{jet} and cores \cite{stecore}, in GRB's \cite{waxman} and in TD 
scenarios using highest injection rates allowed in ref. \cite{prostan}, 
illustrating uncertainties. Recent bounds 
for mechanisms that produce neutrinos trough proton interactions with 
photon fields (such as AGN jet models or GRB's), obtained by demanding 
that cosmic rays are not overproduced, are also shown  
in the extremes of an optically thin target \cite{waxman} and a target 
which is optically thick to neutrons \cite{protmann}. 

Expected neutrino fluxes above the PeV are low and for detection 
large natural target volumes need to be instrumented. Conventional 
detectors use photomultipliers in water or ice to detect the 
Cherenkov light emitted by the long range muons produced in 
charged current muon neutrino interactions, which retain the 
neutrino direction to about a degree. Upcoming events are 
exclusively due to interactions of neutrinos that have travelled 
through the Earth. For neutrinos well above $\sim$ PeV when 
the Earth is opaque, only downgoing to horizontal neutrino events 
are expected which must be separated from the atmospheric muons
\footnote{This is not the case for $\nu_\tau$, see ref.\cite{saltzberg}}.  
The Cherenkov light from high energy showers may allow such 
separation provided the detector has reconstruction capabilities. 
Detection of neutrino induced showers is moreover sensitive to all 
flavor neutrino interactions. These detectors are 
in good development and are likely to bring information on neutrino 
fluxes in very short time \cite{TASI}. 

Alternatively the atmosphere can be instrumented to detect 
horizontal (or upcoming) showers. At sea level the atmosphere is 
roughly 36 times deeper in the horizontal than in the vertical 
direction. As a result horizontal showers induced by comic rays 
get absorbed and only the shower muons can be detected at sea level. 
Showers induced by cosmic rays in the horizontal direction differ from 
typical ones (vertical) in that they hardly have any electron 
or photon density, the arrival time of particles on the ground 
has less spread and agrees better with that of an ideal plane of 
simultaneous muons and lastly they have a characteristic double 
ellipse density profile due to the magnetic field of the Earth.  
It is known since the 60's that penetrating particles such 
as muons and neutrinos can induce showers in the horizontal 
direction \cite{berez} that look like typical 
vertical showers. At very high energies the atmospheric muon 
flux becomes negligible and only neutrinos can produce deep 
showers which can be 
identified provided the detector has an adequate rejection 
power for those unusual ''muon showers'' induced by cosmic rays. 

The recently approved Pierre Auger Observatory in Argentina, 
will be the largest installation (3000 km$^2$) to measure 
air showers. Such a detector array (plans are to build two, 
one in each Hemisphere) can do this separation at least 
based on both timing and muon content and has an acceptance 
of order 10 km$^3$~sr water equivalent \cite{Cronin}. The rate 
of cosmic ray background showers has been estimated to be a few 
thousand showers per year depending on trigger conditions 
\cite{Billoir}.   

As a second alternative Antarctic ice can be instrumented 
with radio antennas to detect the coherent radio pulses 
produced by high energy showers. The idea dates also from 
the 60's and it is particularly attractive for high energy 
showers. Provided the wavelength is larger than the relevant 
shower dimensions the emission from all shower particles is 
coherent and that from electrons and positrons cancels out. 
But as matter electrons constitute the target for the dominant 
interactions below the critical energy ($E_c \sim 73$~MeV in ice), 
an excess negative charge develops in the shower averaging to 
about $20 \%$ of the shower size. As a result the electric field
becomes proportional with the excess charge and since this scales with
shower energy, the power in radio 
emission becomes proportional to the square of shower energy. 
Many experimental difficulties are however anticipated \cite{jelley} 
and presently antennas are being tested deep under ice using the 
AMANDA bore holes in Antarctica (RICE) \cite{rice}.

The first numerical calculation of the radio pulse frequency 
spectrum in the Fraunhofer approximation for electromagnetic 
showers in ice up to 10 PeV revealed a rich 
diffractive pattern but established a threshold of about 
10~PeV for detection at distances above 1~km. It suggested 
that the technique could become competitive for higher energies. 
The simulation becomes increasingly problematic from the 
computational point of view for energies roughly above 
10 PeV as particles have to be tracked down to kinetic energies in the 
100 keV range. Alternative calculation methods are crucial if 
this possibility is ever to be seriously considered. The 
simulation of EeV showers in ice has only been recently 
approached both for electromagnetic and hadronic type showers 
with a method combining simulation and parameterizations in 
the one dimensional approximation. This is also the first time that 
the Cherenkov light in a dense medium is studied in this 
energy range. 

\begin{figure}[hbt]
\centering
\mbox{\epsfig{figure=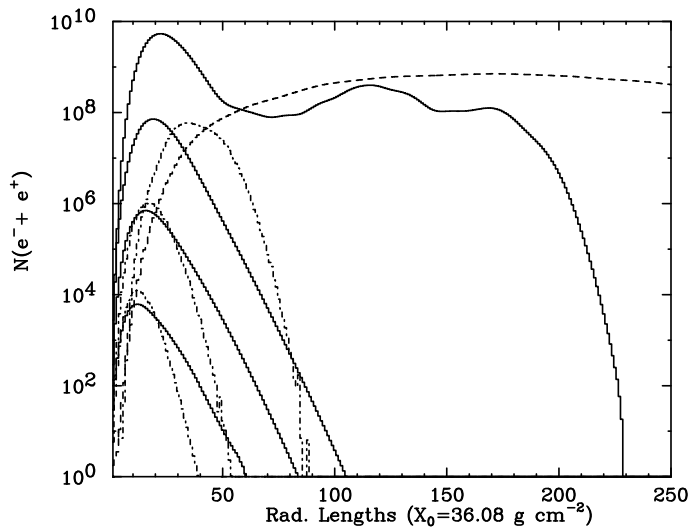,height=7.0cm}}
\caption{Longitudinal development of electromagnetic (dashed curves) and 
hadronic (solid curves) showers in ice. The energies shown are from
bottom to top 10 TeV, 1 PeV, 100 PeV and 10 EeV.}
\label{longemhad}
\end{figure}

For energies below about 20 PeV in ice showers induced by 
electrons or photons produce very similar pulses to those 
induced by hadrons. 
At very high energies however showers behave very 
differently in dense media because of the Landau-Pomeran\v cuck-
Migdal (LPM) effect. As the incident photon or electron rises 
its energy the characteristic length of interactions with an 
electrostatic potential rises to become larger than the 
interatomic spacing. Collective atomic effects manifest as 
a drastic reduction of the cross section for pair production 
and bremsstrahlung that govern shower development. The LPM effect
also suppresses the central part of the differential cross section
for pair production (where the electron and positron carry similar
fractions of the incoming photon energy) and cuts the cross section 
off for bremsstrahlung of low energy photons. As a result 
the showers can become very long (hundreds of radiation lengths).  
Hadronic showers show much smaller elongation because most of 
shower electrons and photons come from decays of $\pi^0$ produced 
in hadronic interactions. Pion decay in ice above 40 PeV is suppressed 
because they are more likely to interact so that even for EeV 
showers only a small fraction of the shower is subject to LPM 
elongations \cite{Zal97,Zal98}. In Fig. \ref{longemhad} the 
simulation results for the developments of hadronic and 
electromagnetic showers in ice are compared. 

\begin{figure}[hbt]
\centering
\mbox{\epsfig{figure=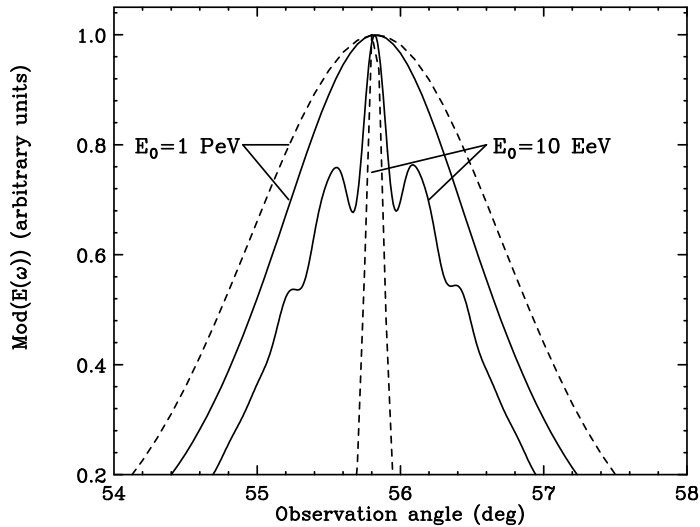,height=7.0cm}}
\caption{Angular distribution of radiopulse around the Cherenkov angle
for the electromagnetic (dashes) and hadronic (solid) showers shown 
in Fig. \ref{longemhad} for shower energies 1 PeV and 10 EeV.}
\label{radioang}
\end{figure}

The LPM has implications 
for any detector sensitive to the showers and  
in principle allows the separation of showers induced by 
electrons in charged current electron neutrino interactions 
from the other showers which are produced by 
the nucleon fragments \cite{Zal99}. The radio emission from
a shower can be viewed, as a first approximation, as the Fourier 
transform of the longitudinal development of the shower. The modifications
introduced by the LPM effect are 
quite dramatic as the angular spread of the diffraction 
pattern narrows linearly as the shower elongates \cite{Zal97}. 
For hadronic showers the pattern shows two angular periodicities 
corresponding to the two shower scales \cite{Zal98} see 
Fig. \ref{radioang}. 

In summary the confirmation of cosmic rays of energies above $10^{20}$~eV 
is very suggestive for the existence of EeV neutrino fluxes at levels 
which may be detectable in the foreseeable future. Conventional neutrino detectors will have the first word on high energy neutrino fluxes but 
alternative techniques can also contribute particularly in the EeV 
region. While horizontal shower measurements will play an important 
role as the next generation of detectors has an acceptance which is 
of order 10~km$^3$sr, the radio technique provides a most interesting 
possibility which may turn out to be most adequate if acceptances 
above km$^3$ are required. They moreover have many added advantages 
due to the coherence character of the signal to be measured which will 
be of great use in trying to establish neutrino flavor, particularly 
if combined with other techniques.    

\section*{Acknowledgments}
We are indebted to M\'aximo Ave, Gonzalo Parente 
and Ricardo V\'azquez who have collaborated with work reported 
here and we thank CICYT (AEN96-1773) and Xunta de Galicia (XUGA-20602B98) 
for supporting this research. J. A. thanks the Xunta de Galicia for financial
support.


\section*{References}
\begin{thebibliography}{99}
%
\bibitem{solar} J.~Bahcall, in {\sl Proc. of the 18th Texas Symposium on Relativistic Astrophysics} eds. A. Olinto, J. Frieman and D.Schramm (World Scientific, Singapore, 1998) pp. 99-113. e-Print Archive: astro-ph/9702057.
%
\bibitem{sn1987A} See for instance D.N.~Schramm, 
{\sl Comments Nucl. Part. Phys.} {\bf 17} 5 (1987) pp 239-278 and references therein.
%
\bibitem{kamioka} K.S.~Hirata {\sl et al.}, {\it Phys. Lett.} {\bf B280} (1992) 146-152.
%
\bibitem{SNnumass} L.F.~Abbott {\sl et al.} {\sl Nucl. Phys.} {\bf B299} (1988) 734-756.
%
\bibitem{oscillations} Y. Fukuda {\it et al.}, Phys. Rev. Lett. {\bf 81}
(1998) 1562.
%
\bibitem{stecore}
F.W.~Stecker, C.~Done, M.H.~Salamon and P.~Sommers, {\it Phys.\ Rev.\ Lett.}
{\bf 69}, 2738(E) (1992).
%
\bibitem{jet} K. Mannheim, Astrop. Phys. 3 (1995) 295-302; 
R.J. Protheroe, in Proc. IAU Colloq. 163, Accretion
Phenomena and Related Outflows, ed. D. Wickramasinghe et al., in
press (1996).
%
\bibitem{waxmangrb} E. Waxman and J. Bahcall, {\it Phys.\ Rev.\ Lett.} 
{\bf 78} (1997) 2292.
%
\bibitem{TD}  P.~Battacharjee, C.T.~Hill, D.N.~Schramm, Phys.\ Rev.\
Lett.\ {\bf 69} (1992) 567.  
%
\bibitem{physrep} T.K. Gaisser, F. Halzen and T. Stanev, Phys. Rep.
238 (1995) 173.
%
\bibitem{HECR} K. Greisen, {\it Phys. Rev. Lett.}, {\bf 16}, 748 (1966); 
G.T. Zatsepin and V.A.  Kuz'min, {\it JEPT Lett.}, {\bf 4}, 78, (1966).
%
\bibitem{weiler} T.K. Weiler, Astrop. Phys. (in press), e-Print Archive: hep-ph/9710431; E. Waxman e-Print Archive: astro-ph/9804023. 
%
\bibitem{Volkova} L.V.~Volkova {\sl et al.}, {\sl Nuovo\ Cim.\ } {\bf 10C}  
(1987) 465. 
%
\bibitem{domokos} G.~Domokos {\sl et al.}, {\sl J. Phys. G: Nucl. Part. Phys.} 
{\bf 19} (1993) 899. 
%
\bibitem{nuGZK}  S. Yoshida and M. Teshima, Prog. of Theo. Phys. 89 (1993) 833;  
F.W.~Stecker, C.~Done, M.H.~Salamon and P.~Sommers,
{\sl Phys.\ Rev.\ Lett.\ } {\bf 66} (1991) 2697.
%
\bibitem{sigl} S. Yoshida, G. Sigl and  S. Lee; {\it Phys.\ Rev.\ Lett.}  {\bf 58} (1998) 5505-5508. 
%
\bibitem{waxman} E. Waxman and J. Bahcall, {\it Phys. Rev.} {\bf D59} (1999) 023002.
%
\bibitem{prostan} R.J. Protheroe and T. Stanev, {\it Phys. Rev. Lett.}, {\bf
77}, 3708 (1996), and {\it Erratum} {\bf 78}, 3420 (1997).
%
\bibitem{protmann} K. Mannheim, R. J. Protheroe and J. P. Rachen, e-print 
Archive: astro-ph/9812398
%
\bibitem{saltzberg} F. Halzen and D. Saltzberg, {\sl Phys. Rev. Lett.} {\bf 81}
(1998), 4305-4308.
%
\bibitem{TASI} F. Halzen; {\it Lectures on Neutrino Astronomy: Theory and Experiment}
TASI School, July 1998; e-print Archive astro-ph/9810368.
%
\bibitem{berez} V.S.~Berezinsky and G.T.~Zatsepin {\sl Phys.\ Lett.\ }
{\bf B28} (1969) 423; P. Kiraly {\it et al.}, {\sl J. Phys. A:Gen. Phys.} {\bf 4} 
(1971) 367. V.S. Berezinsky and A. Yu. Smirnov, {\sl Astrophys. Space
Science} {\bf 32} (1975) 461.
%
\bibitem{Cronin} J.~Capelle, J.W.~Cronin, G.~Parente and E.~Zas;
{\sl Astropart. Phys.} {\bf 8} (1998) 321.
%
\bibitem{Billoir} M.~Ave 1998 diploma thesis unpublihsed; P.~Billoir 
private comm.
%
\bibitem{jelley} J.V.~Jelley. {\sl Astropart. Phys.} {\bf 5} (1996)255-261. 
%
\bibitem{rice} C.~Allen {\it et al.}, {\sl Proc.\ XXVth ICRC}, Durban, South Africa, Ed. M.S.~Potgieter {\it et al.} (1997), Vol.~7, pp. 85--88.
%
\bibitem{Zal97} J.~Alvarez-Mu\~niz and E.~Zas, {\sl Phys.\ Lett.} 
{\bf B411} (1997) 218-224. 
%
\bibitem{Zal98} J.~Alvarez-Mu\~niz and E.~Zas,
{\sl Phys.\ Lett.} 
{\bf B434} (1998) 396-406. 
%
\bibitem{Zal99} J.~Alvarez-Mu\~niz, R.A.~Vazquez and E.~Zas, 
e-print Archive: astro-ph/9901278 (1999) . 
%
\end{thebibliography}
\end{document}